\documentclass{iaus}
\usepackage{graphicx}

\title[GRBs as probes of massive stars near and far] %% give here short title %%
{GRBs as probes of massive stars near and far}

\author[Johan P. U. Fynbo]   %% give here short author list %%
{Johan P. U. Fynbo$^1$,
Daniele Malesani$^1$}

\affiliation{$^1$Dark Cosmology Centre, Niels Bohr Institute, University of Copenhagen, \\ DK-2100 Copenhagen O,
Denmark \\ email: {\tt jfynbo@dark-cosmology.dk}%\\[\affilskip]
}

\pubyear{2008}
\volume{250}  %% insert here IAU Symposium No.
\pagerange{1-13}
\jname{Massive stars as cosmic engines}
\editors{F. Bresolin, P. A. Crowther \& J. Puls, eds.}
\begin{document}

\maketitle

\begin{abstract}
Long-duration gamma-ray bursts are the manifestations of massive stellar death.
Due to the immense energy release they are detectable from most of the
observable universe. In this way they allow us to study the deaths of single
(or binary) massive stars possibly throughout the full timespan massive stars
have existed in the Universe. GRBs provide a means to infer information about
the environments and typical galaxies in which massive stars are formed. Two
main obstacles remain to be crossed before the full potential of GRBs as probes
of massive stars can be harvested: {\it i}) we need to build more complete and
well understood samples in order not to be fooled by biases, and {\it ii)} we
need to understand to which extent GRBs may be intrinsically biased in the
sense that they are only formed by a limited subset of massive stars defined by
most likely a restricted metallicity interval. We describe the status of an
ongoing effort to build a more complete sample of long-duration GRBs with
measured redshifts. Already now we can conclude that the environments 
of GRB progenitors are very diverse with metallicities ranging from solar to
a hundredth solar and extinction ranging from none to $A_V>5$ mag. We have also 
identified a sightline with significant escape of Lyman continuum photons
and another with a clear 2175 \AA \ extinction bump.
\end{abstract}

\firstsection % if your document starts with a section,

\section{Introduction}

GRBs were discovered serendipitously in the late 1960ies and first reported to
the astrophysical community in the early 1970ies (\cite[Klebesadel et al.
1973]{klebesadel73}). For a long time their nature remained a mystery
(\cite[e.g., Nemiroff 1994]{nemiroff93}). For a review of the first decades of
GRB research where only the prompt $\gamma$-ray emission was known see Fishman
\& Meegan (1995). The major breakthrough came in 1997 with the ability to
determine celestial positions of GRBs thanks to the BeppoSax satellite and the
discovery of long-lived X-ray, optical and radio afterglows (\cite[e.g., Costa
et al. 1997; van Paradijs et al. 1997; Frail et al. 1997] {costa97,
vanParadijs97,frail97}). For a review
of the early years of the so called afterglow era (after 1997) see van Paradijs
et al. (2000). 

GRBs come in at least two variants defined by their duration in the
$\gamma$-ray band.  The short bursts have duration less than about 2 s and
the long bursts longer than this (\cite[Kouveliotou et al. 1993]{chryssa93}).
In the reminder of this review we will only discuss the long duration bursts. The association
between long-duration GRBs (LGRBs hereafter) and massive stars and hence the
link between LGRBs and on-going massive star-formation found its first
empirical basis with the detection of the first host galaxies (e.g., \cite[Hogg
et al. 1999]{hogg99}). Subsequently, the evidence was further strengthened with
the discovery of supernovae (SNe) associated with LGRBs (\cite[Galama et al.
1998; Hjorth et al. 2003; Stanek et al. 2003; Malesani et al. 2004; Sollerman
et al. 2006; Pian et al.
2006]{galama98,hjorth03,stanek03,malesani04,sollerman06,pian06}). For a recent
review of the LGRB/SN association see \cite[Woosley \& Bloom (2006)]{woosley06}.

Because of the link between LGRBs and massive stars and due to the fact that
GRBs can be detected from both the most distant and the most dust obscured
regions in the universe LGRBs were quickly identified to be very promising
tracers of star-formation throughout cosmic history (e.g., \cite[Wijers et al.
1998]{wijers98}). However, this potential has so far not really resulted in
an improved census of the locations of massive stars due to complications
discussed in the next section.

A major issue currently under discussion is if LGRBs are unbiased tracers of
star formation. More precisely, it is not clear if LGRBs are caused by the same
(small) fraction of all dying massive stars (unbiased tracers), or if LGRBs only
trace a limited segment defined by parameters such as, e.g., metallicity or
circumstellar density (biased tracers).

The currently operating {\it Swift} satellite \cite[(Gehrels et al. 2004)]{gehrels04} has
revolutionized LGRB research with its frequent, rapid, and precise localization
of LGRBs. Now it is for the first time possible in practice to use LGRBs as
powerful probes. It is mandatory that this potential is exploited while {\it
Swift} is still operating (at least until 2010).

\section{Complications in the use of LGRBs as tracers of massive stars}

\subsection{Dark bursts and incomplete samples}
A crucial issue when using LGRBs (or any other class of tracer) 
is sample selection. Whereas the detection of the LGRB itself
poses no bias against dust obscured massive stars this is not the case for the
softer afterglow emission which is crucial for obtaining the precise
localization as well as measuring redshifts (see, e.g., 
\cite[Fiore et al. 2007]{fiore07}).

In the samples of LGRBs detected with satellites prior to the currently
operating {\it Swift} satellite the fraction of LGRBs with detected optical
afterglows was only about 30\% \cite[(Fynbo et al. 2001; Lazzati et al.
2002)]{fynbo01,lazzati02}.  Much of this incompleteness was caused by random
factors such as weather or unfortunate celestial positions of the bursts, but
some remained undetected despite both early and deep limits. It is possible
that some of these so called ``dark bursts'' could be caused by LGRBs in very
dusty environments \cite[(Groot et al. 1998)]{groot98} and hence the sample of
LGRBs with detected optical afterglows could very well be systematically biased
against dust obscured star formation (see also \cite[Jakobsson et al. 2004a;
Rol et al. 2005; Rol et al. 2007]{palli04a,rol05,rol07} for
recent discussions of the dark bursts).

In any case, such a high incompleteness imposes a large uncertainty on
statistical studies based on LGRB host galaxies derived from these early
missions. It should be stressed that the conclusions based on these 
samples may only be relevant for a minority of all LGRBs.
Due to the much more precise and rapid localization capability of {\it
Swift} it is now possible to build much more complete samples. 

\subsection{Are some LGRBs not associated with massive stellar death?}

Recently, it has been found that some LGRBs are not associated with SNe, namely
GRB\,060505 \cite[(Fynbo et al. 2006a; Ofek et al. 2007)]{fynbo06a,ofek07} and
GRB\,060614 \cite[(Fynbo et al. 2006a; Della Valle et al. 2006; Gal-Yam et al.
2006)]{fynbo06a,dellavalle06,gal-yam06}. This means that either some massive
stars die without producing SNe brighter than about $M_\mathrm{V}$=-13.5 
(\cite{fynbo06a, dellavalle06}) or, alternatively, some LGRBs are caused by
other mechanisms than collapsing massive stars \cite[(Gal-Yam et al. 2006;
Ofek et al. 2007)]{gal-yam06,ofek07}.

\begin{figure}[h]
% \vspace*{-2.0 cm}
\begin{center}
 \includegraphics[width=5.3in]{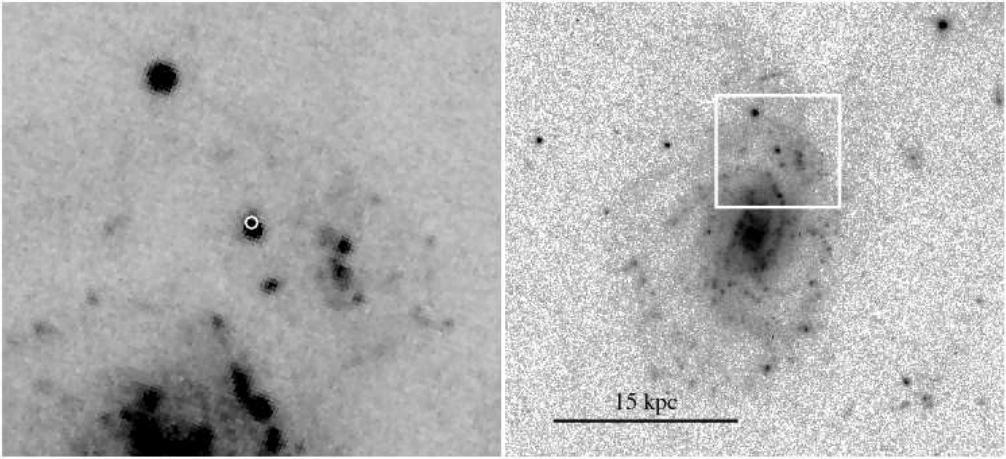} 
% \vspace*{-1.0 cm}
 \caption{
The host galaxy of GRB\,060505 as observed with the HST \cite[(Ofek et al.
2007; Th\"one et al. 2008)]{thoene08,ofek07}. The white circle shows the
error-circle of the burst consistent with a star-forming region in a spiral
arm. This 
is strong evidence that the progenitor was a massive star.  The properties of
this host is also within the range found for other LGRBs. As an example,
the host galaxy of GRB\,050824 is very similar in terms of luminosity, R23 and
star-formation rate \cite[(Sollerman et al. 2007)]{sollerman07}, and the
location within the host is similar to the location of other LGRBs in spiral
hosts \cite[(Fynbo et al. 2000a; Sollerman et al. 2002; Le Floc'h et al. 2002;
Jakobsson et al. 2005a)]{fynbo00a,sollerman02,lefloch02,palli05a}.
 }
\label{fig1}
\end{center}
\end{figure}

At least for the case of GRB\,060505 the evidence points to the former.  The
burst was located in a star-forming region in a relatively low metallicity
region in the outer part of a spiral host (\cite[Fynbo et al. 2006a; Ofek et al.
2007; Th\"one et al. 2008]{fynbo06a, ofek07,thoene08}; see also
Fig.\,\ref{fig1}). The burst itself displayed a significant spectral lag, which
has so far never been seen for the short bursts that are believed to originate
from merging compact objects (\cite[McBreen et al. 2008; Norris \& Bonnell
2006]{mcbreen08,norris06}). If some LGRBs indeed are caused by other
progenitors than massive stars then a new classification that can distinguish
between LGRBs from massive stars and those from other mechanisms is required.
So far no such scheme has been found (see \cite[Gehrels et al.
2006]{gehrels06}).

\subsection{The contamination from chance projections}
\begin{figure}
\begin{center}
 \includegraphics[width=5in]{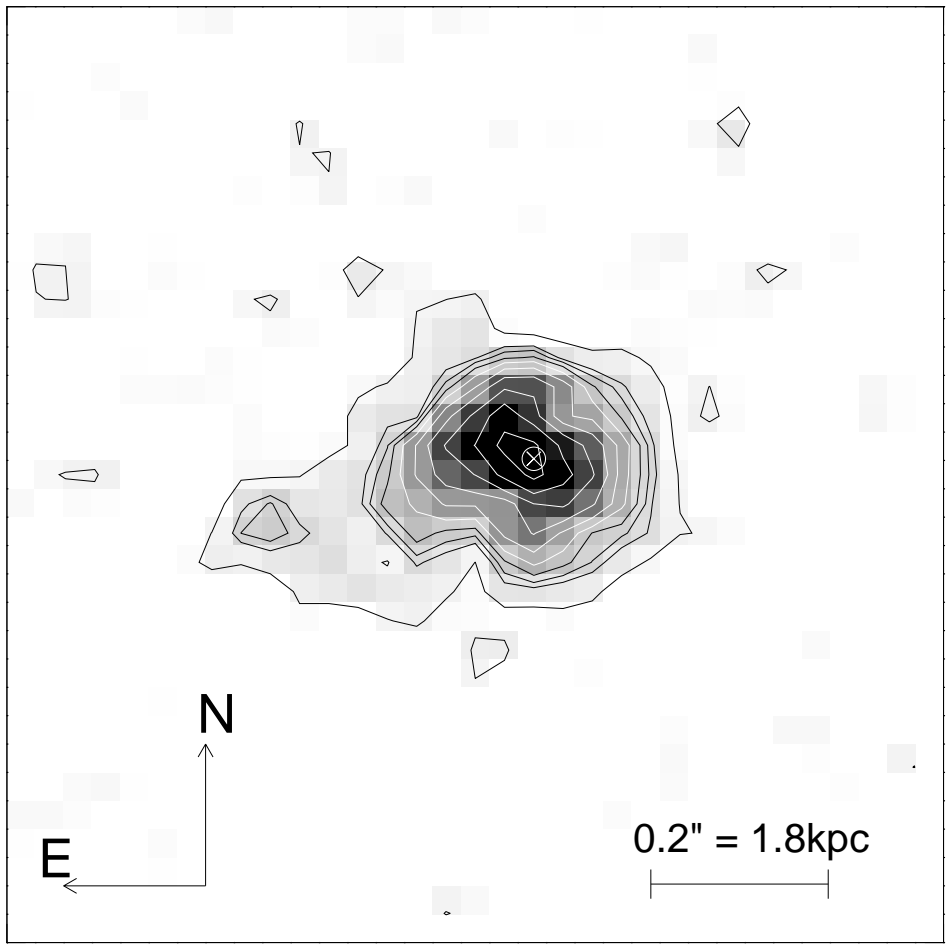} 
\caption{
The 1$\times$1 arcsec$^2$ field around the host galaxy of the $z=2.33$ HETE-2
GRB\,021004 observed with the {\it HST} (from \cite[Fynbo et al.
2005]{fynbo05}). The LGRB went off near the center of the galaxy. The position
of the LGRB is marked with a cross and an error circle and it coincides with the
centroid of the galaxy to within a few hundredths of an arcsec.  In cases like
this there is no problem in identifying the correct host galaxy. However, in
cases of bursts localized to only a few arcsec accuracy chance projection
needs to be considered.
}
\label{021004host}
\end{center}
\end{figure}

The first important question to ask is: are LGRB host galaxies operationally
well-defined as a class? In terms of an operational definition the case is not
so clear. If we define the host galaxy of a particular burst to be the galaxy
nearest to the line-of-sight, we need to worry about chance projection
\cite[(Band \& Hartmann 1998)]{band98}. In the majority of cases where an
optical afterglow has been detected and localized with subarcsecond accuracy
and where the field has been observed to deep limits a galaxy has been detected
within an impact parameter less than 1 arcsec (see e.g., \cite[Bloom et al.
2002]{bloom02} and Fig.\,\ref{021004host} for an example).  The probability for
this to happen by chance depends on the magnitude of the galaxy. The number of
galaxies per arcmin$^2$ has been well determined to deep limits in the Hubble
deep fields.  In \cite[Fynbo et al. (2000b, their Fig.~2)]{fynbo00b} the galaxy
counts in the R and I bands based on the HDF-South can be found. To limits of
R=24, 26 and 28 there are about 2, 6 and 13 galaxies arcmin$^{-2}$. Hence, the
probability to find a R=24 galaxy by chance in an error circle with radius 0.5
arcsec is about 4$\times$10$^{-4}$. For a R=28 galaxy the probability is about
3$\times$10$^{-3}$. If the error circle is defined only by the X-ray afterglow
with a radius of 2 arcsec in the best cases then we expect a random R=24 and
R=28 galaxy in 0.6\% and 5\% of the error circles. For a sample of a few
hundred LGRBs chance projection should hence not be a serious concern for LGRBs
localized to sub-arcsecond precision, but for error-circles with radius of a
few arcseconds we expect a few chance projections. In some cases it may be
possible to eliminate the chance projects, e.g., based on conflicting 
redshift information from the afterglow and proposed host, but in general
not.

\section{Conclusions based on pre-Swift samples}

Despite the complications mentioned above, the previous decade of LGRB host
galaxy studies has after all taught us a lot about LGRBs and their link to
massive star formation (see \cite[van Paradijs et al. 2000 and Djorgovski
2003]{vanparadijs00,djorgovski03} for early reviews). LGRB hosts were early
on found to be predominantly faint, blue star-forming galaxies \cite[(Hogg et
al. 1999)]{hogg99}. The early studies found that these properties of LGRB hosts
were consistent with the expectation if LGRBs are unbiased tracers of
star-formation \cite[(Mao et al. 1998; Hogg et al. 1999)]{mao98,hogg99}. It
was also realized early on that LGRBs offer a unique possibility to locate and
study star-formation activity in dwarf galaxies at $z>2$ \cite[(Jensen et al.
2001)]{jensen01}. This is basically impossible with any other currently
existing method. The star-formation rates were found to be modest, but the
specific star-formation rates among the highest ever found
\cite[(Christensen et al. 2004)]{christensen04}. LGRB hosts are hence often 
in a starburst state. 

Later evidence indicated that LGRBs maybe related only to massive stars
with metallicity below a certain threshold. The first evidence for this came
with the realization the LGRB hosts were fainter and bluer than expected
according to certain models about the nature of the galaxies dominating the
integrated star-formation activity \cite[(Le Floc'h et al.  2003, 2006; Tanvir
et al. 2004)]{lefloch03,lefloch06, tanvir04}. Nevertheless, it has recently been
pointed out by \cite[Priddey et al. (2006)]{priddey06} that ``there is
sufficient uncertainty in models and underlying assumptions, as yet poorly
constrained by observation (e.g., the adopted dust temperature) that a
correlation between massive, dust-enshrouded star formation and GRB production
cannot be firmly ruled out.'' (see also \cite[Micha\l{}owski et al. 2008 concerning
the issue of dust temperature]{michal08}). 
Further circumstantial evidence for a preference
towards low metallicity came from the observation that Lyman-$\alpha$ emission
seemed to be ubiquitous for LGRBs hosts \cite[(Fynbo et al. 2003; Jakobsson et
al. 2005b)]{fynbo03,palli05b}. 

Lately, evidence from hosts of more local LGRBs seems to point in the same
direction. Several studies have found that local LGRBs hosts tend to be faint
and metal poor although a caveat for some of these studies is the difficulty of
using the strong line metallicity indicators like R23 to derive robust
metallicities \cite[(Prochaska et al. 2004; Sollerman et al. 2005; Gorosabel et
al. 2005; Stanek et al. 2006; Wiersema et al.
2007)]{prochaska04,sollerman05,gorosabel05,stanek06,wiersema07}.  However,
nearly all of these studies targeted very incomplete pre-{\it Swift} samples
and this raises the question whether the predominance of faint, metal poor hosts
can be explained by, e.g., a bias against metal rich hosts and hence more dust
obscured LGRB afterglows.

A very important result is that LGRBs and core-collapse SNe are found in
different environments \cite[(Fruchter et al. 2006)]{fruchter06}. The same
study also found that LGRB host galaxies at $z<1$ are fainter than the host
galaxies of core-collapse SNe.  This study is also based on incomplete pre-{\it
Swift} samples, but as the SNe samples are if anything more biased against
dusty regions than LGRBs this result does seem to be substantial evidence that
LGRBs are biased towards massive stars with relatively low metallicity.
\cite[Wolf \& Podsiadlowski (2007)]{wolf07} however, find, based on an analysis
of the Fruchter et al. (2006) 
data, that the metallicity threshold cannot be below half the solar
metallicity. Concerning the different environments of core-collapse SNe and
LGRBs it has recently been found that type Ic SNe have similar positions
relative to their host galaxy light profiles as LGRBs, whereas all other SN
types have a similar distribution, less centred on their host light than LGRBs
and SN Ic's \cite[(Kelly et al. 2007)]{kelly07}. \cite[Larsson et al.
(2007)]{larsson07} find that the different distributions of different SN types
relative to their host light can be naturally explained by assuming different
mass ranges for the typical progenitor stars: $\gtrsim$8 M$_{\odot}$ for
typical core-collapse SNe and $\gtrsim$ 20 M$_{\odot}$ for LGRB progenitors.
The picture is complicated by the finding that type Ic SNe typically are found
in substantially more metal rich environments than LGRBs \cite[(Modjaz et al.
2008, and in these proceedings)]{modjaz08}. It is well established that
WR-stars become more abundant with {\it increasing} metallicity - opposite to
LGRBs that if anything are biased towards low metallicity. Taken together these
findings suggest that progenitors of LGRBs and ``normal'' type Ic SNe are two
different subsets of the $\gtrsim$ 20 M$_{\odot}$ stars. For a thorough
discussion of the relation between WR stars, SN Ic's and LGRBs we refer to
\cite[Crowther (2007)]{crowther07}.

\section{Building a complete sample of {\it Swift} LGRBs}

The following is to a large extent based on Fynbo et al. (2007).
The {\it Swift} satellite has been operating for about three years and is far
superior to previous GRB missions. The reason for this is the combination of
several factors: {\it i)} it detects LGRBs at a rate of about two bursts per
week about an order of magnitude larger than the previous successful BeppoSAX
and HETE-2 missions; {\it ii)} with its X-Ray Telescope (XRT) it localizes the
bursts with a precision of about 5 arcsec  also orders of magnitude better than
previous missions; {\it iii)} it has a much shorter reaction time, allowing the
study of the evolution of the afterglows literally seconds after the burst,
sometimes during the prompt $\gamma$-ray emission itself. The {\it Swift}
mission is funded at least until 2010. A crucial objective is to secure a large
sample, as complete as possible, of LGRB afterglows while {\it Swift} is still
operating. More concretely, rather than including all {\it Swift} detected
GRBs, it is more optimal to concentrate on those LGRB afterglows with 
favourable observing conditions. Our group uses the the following sample
criteria: 
\begin{enumerate}
\item{XRT afterglow detected within 12 hr}
\item{Small foreground Galactic extinction: $A_V<0.5$ mag}
\item{Favourable declination: $70 < dec < 70$}
\item{Sun distance larger than 55$^o$}
\end{enumerate}
By introducing these constraints, we
are not biasing the sample towards optically bright afterglows, but we select a
sample for which useful follow-up observations are likely to be secured. 

About 50\% of all {\it Swift} LGRBs do not fulfill these criteria, primarily
because Swift, for technical reasons, has to point close to the Sun a
significant fraction of the time. For bursts fulfilling the above criteria, we
make every possible effort to detect optical and near-infrared afterglows and to
measure their redshifts.  As shown below, we have been very successful in this
effort, using mainly the ESO VLT. Redshifts, or more generally spectroscopic
observations, are crucial for almost all LGRB-related science. The most
important science cases for which spectroscopy is critical are listed below:
\begin{itemize}
\item{Determining the luminosity function for LGRBs (prompt emission as well as
afterglows)}
\item{Determining the redshift distribution of LGRBs and using LGRBs as tracers
for the cosmic star-formation history (Jakobsson et al. 2006a; Fiore et al.
2007)}
\item{Studying the host galaxies, in particular those faint, high-redshift
galaxies that are unlikely to be found and studied with other methods (e.g.,
Vreeswijk et al. 2004)}
\item{Studying LGRB-selected absorption-line systems (e.g., Jakobsson et al. 2004b; Prochter et al.  2006)} 
\item{Characterizing the dust extinction curves of high-z galaxies (e.g., 
Jensen et al. 2001; see also Fig.~\ref{bump})}
\item{Determining the Lyman continuum escape fraction from high-z galaxies (Chen
et al. 2007, see also Fig.~\ref{LyC})}
\item{Spotting very high redshift LGRBs (e.g., Kawai et al. 2006; 
Ruiz-Velasco et al. 2007)}
\item{Probing cosmic chemical evolution with LGRBs (e.g., Savaglio 2006; 
Fynbo et al. 2006b; Prochaska et al. 2007)}
\item{Studying if LGRBs can be used for cosmography (e.g., Ghirlanda et al. 
2004)}
\end{itemize}

\begin{figure}[h]
\begin{center}
\includegraphics[width=5.3in]{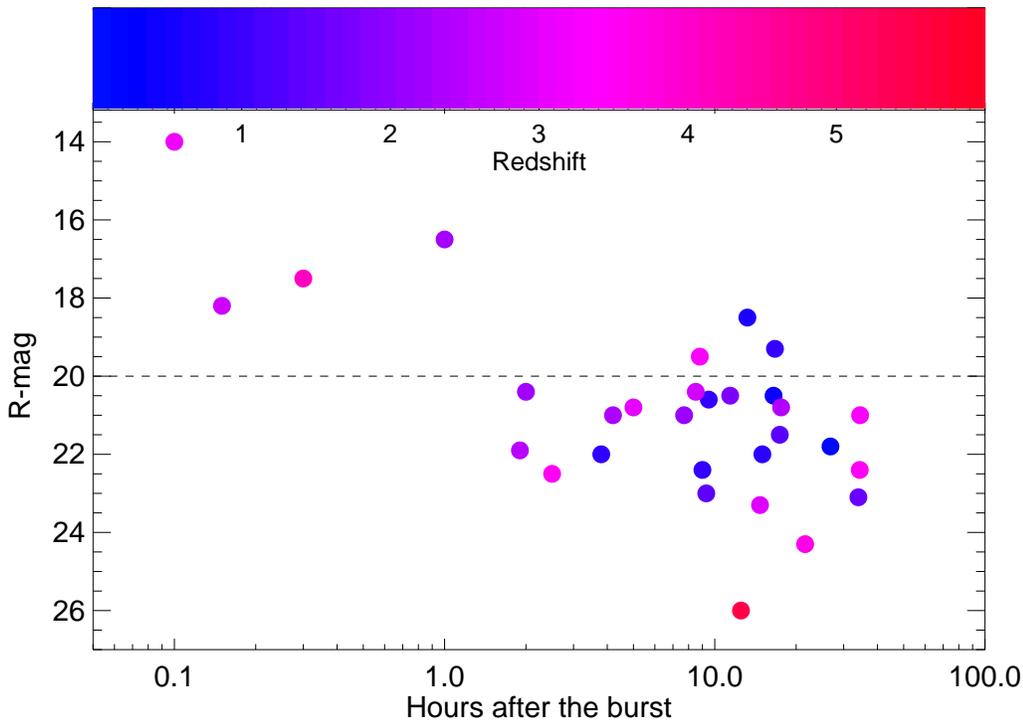} 
\caption{
The R-band magnitude of the optical afterglows as a function of the time after
the burst at which the spectroscopic observations were obtained.  Only included
are Swift bursts for which we have measured the redshift (using primarily the
VLT, but also NOT, WHT and GEMINI). The colour bar at the top indicates the
colour code for the measured redshifts. The dashed line marks a magnitude of $R
= 20$ which is roughly the spectroscopic limit for 2--4-m telescopes for
detecting absorption lines. As seen, most afterglows are fainter than this
limit when observable.
}
\label{dtmag}
\end{center}
\end{figure}

\begin{figure}
\begin{center}
 \includegraphics[width=5in]{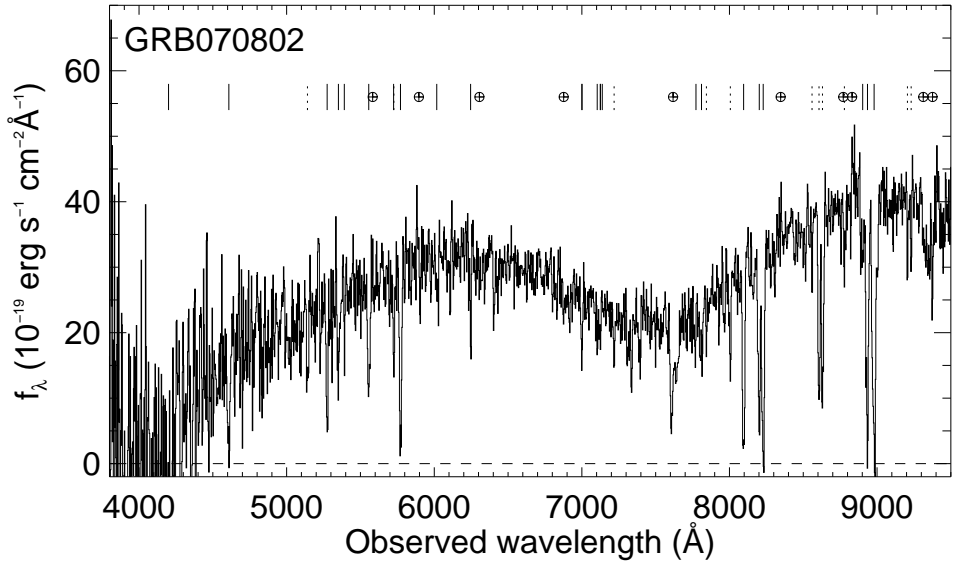} 
\caption{
The VLT/FORS2 spectrum of the afterglow of GRB\,070802 (El\'iasd\'ottir
et al., in preparation). 
Plotted is the flux-calibrated spectrum
against observed wavelength. Metal lines at the host redshift are
marked with full-drawn lines whereas the lines from two intervening systems
are marked with dotted lines. The broad depression centred around 7500 \AA \
is caused by the 2175 \AA \ extinction bump in the host system
at $z_{\rm abs}=2.4549$.
}
\label{bump}
\end{center}
\end{figure}

\begin{figure}
\begin{center}
 \includegraphics[width=5in]{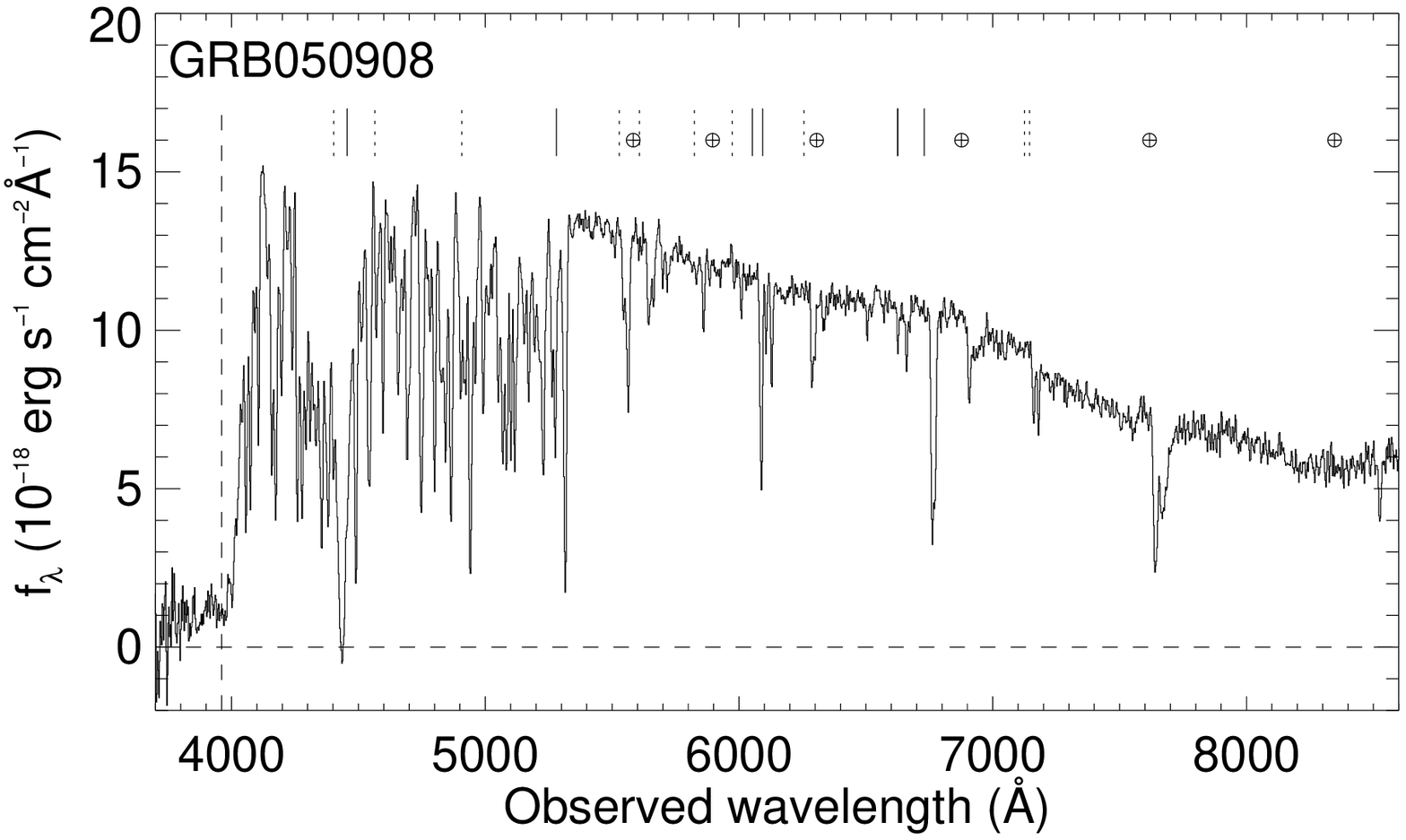} 
\caption{
The VLT/FORS2 spectrum of the afterglow of GRB\,050908 (Smette
et al., in preparation). Plotted is the flux-calibrated 1-dimensional
spectrum against observed wavelength. The vertical
dashed line shows the position of the lyman limit at the 
redshift of the GRB ($z=3.343$). As seen, there
is clear excess flux below the lyman limit.
}
\label{LyC}
\end{center}
\end{figure}

Since the launch of {\it Swift}, we have had programmes running at the VLT with
the aim of securing redshifts for {\it Swift} LGRBs.  The status at the time of
writing is that 109 {\it Swift} LGRBs fulfilled our selection criteria. For 57
of these a redshift measurement has been secured (see Fig.~\ref{zdist}). The
VLT has been the dominant single contributor in all LGRB redshift measurements,
providing around 40\% of the secure redshifts to date.  The redshift of the
most distant known GRB\,050904 at $z=6.295$ was measured with the SUBARU
telescope (Kawai et al. 2006). Most of the other redshifts have been measured
using other 6--10 m telescopes (Keck, GEMINI, SUBARU, Magellan).  This is
contrary to the expectations prior to the launch of {\it Swift}, where it was
suspected that {\it Swift} itself, or at least 2--4-m telescopes, would be able
to measure most of the redshifts.  However, optical afterglows turned out to be
much fainter at early times than anticipated.  As we show in Fig.~\ref{dtmag},
the majority of the afterglows are fainter than $R = 20$ when a slit can be
placed on them. $R = 20$ is, in our experience, the limit for spectroscopic
redshift determination using 2--4-m telescopes (typically, no more than 1--2 hr
exposure time is available for observing LGRB afterglows).  Several optical
afterglows are already fainter than $R = 22$ a few hours after the bursts.
Hence, 6--10 m telescopes are crucial for securing redshifts for the majority of
Swift LGRBs. 

The fact that in particular the VLT, but also other 6-10 m telescopes, have
made tremendous efforts to secure redshifts means that we now have a much
higher redshift completion than for pre-Swift samples. But it is clear that we
will not get redshifts for all bursts from spectroscopy of the afterglows for
multiple reasons. In about 20-30\% of the triggers we are not able to measure
the redshift either due to lack of lines (probably bursts at redshifts between
1 and 2, see Fig.~\ref{zdist}), bad weather or because the afterglow has faded
too much before it is observable from Paranal. For these bursts our only chance
of measuring the redshift is via spectroscopy of the host galaxy. We have also
pursued this route extensively in an ESO large program (PI Hjorth). This is a
challenging task due to the faintness of these systems, and the analysis of
these data is still ongoing, but we have already determined a number of
redshifts (included in Fig.~\ref{zdist}). 

\subsection{The redshift distribution of Swift LGRBs: current status}
\begin{figure}[h]
\begin{center}
\includegraphics[width=5.3in]{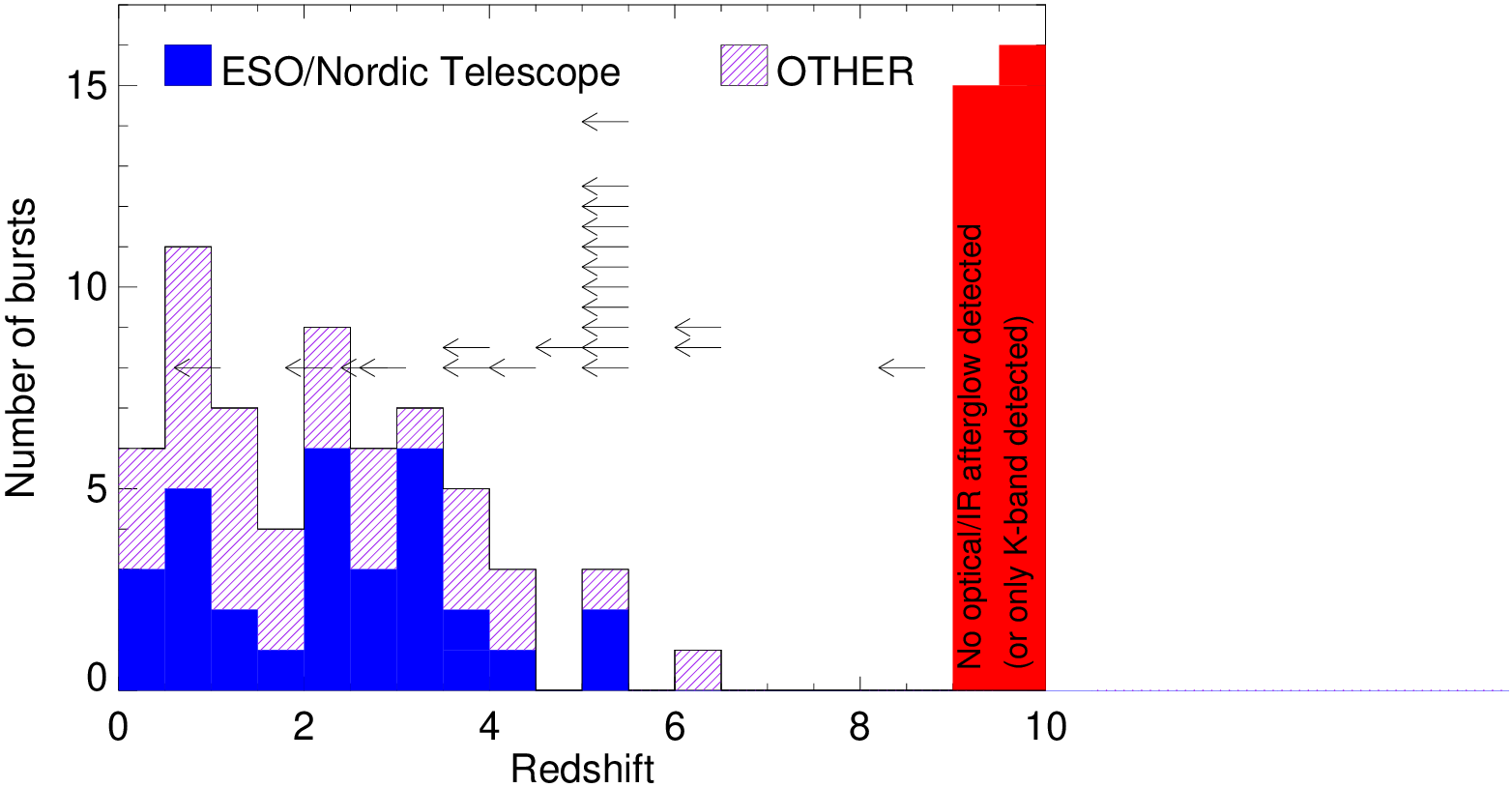} 
\caption{
Redshift distribution (up to October 2007) of 109 Swift LGRBs localized
with the X-ray telescope and with low foreground extinction $A_V<0.5$. Of the 58
measured redshifts, our group has measured nearly half (25, shown in blue). As
shown, the VLT is the dominant source of redshifts in the Swift era (four of
the blue bursts in the histogram are also from the Nordic Optical Telescope).
Bursts, for which only an upper limit on the redshift could be established so
far, are indicated by arrows. Note that it is also difficult to secure
redshifts for GRBs in the desert between $z = 1$ and $z = 2$. The red histogram at
the right indicates the 27 bursts for which no optical/J/H afterglow was
detected and hence no redshift constraint could be inferred (see Ruiz-Velasco
et al. 2007 for a full discussion).
}
\label{zdist}
\end{center}
\end{figure}

The first conclusion from Fig.~\ref{zdist} is that Swift LGRBs are very distant.
Swift LGRBs are more distant than LGRBs from previous missions due to the higher
sensitivity of the satellite to the lower energies prevalent in the more
distant events (Fiore et al. 2007).  The median and mean redshift are now both
2.3, while for previous missions it was closer to 1 (Jakobsson et al. 2006a).
The record holder is $z=6.295$ (Kawai et al. 2006).
It is striking how events at redshifts as large as 6
can be detected within such a small sample. For comparison, only a few QSOs are
detected at similar distances out of a sample of hundred thousand QSOs.
Remarkably, the redshift distribution, measured for just over 50\% of all
bursts, is consistent with the redshift distribution predicted if LGRBs are
unbiased tracers of star formation (see, e.g., Jakobsson et al. 2006a and
http://www.dark-cosmology.dk/$\sim$pallja/GRBsample.html for a regularly updated
analysis). 

\subsection{HI column densities}
The HI column density distribution for LGRB sightlines is extremely broad. It
covers a range of about 5 orders of magnitude from $\sim10^{17}$ cm$^{-2}$
(Fig.~\ref{LyC}, Chen et al. 2007) to nearly $10^{23}$ cm$^{-2}$ 
(Jakobsson et al. 2006b). I still remain to be understood if this distribution
is representative of the intrinsic distribution of HI column densities
towards massive stars in galaxies or if the distribution is rather controlled
by the ionizing emission from the afterglows themselves. In any case, as 
pointed out by Chen et al. (2007) the HI column density distribution provides
an upper limit to the escape fraction of Lyman continuum emission from 
star-forming galaxies.

\subsection{Metallicities} 
\begin{figure}[h]
\begin{center}
\includegraphics[width=4.0in,angle=90]{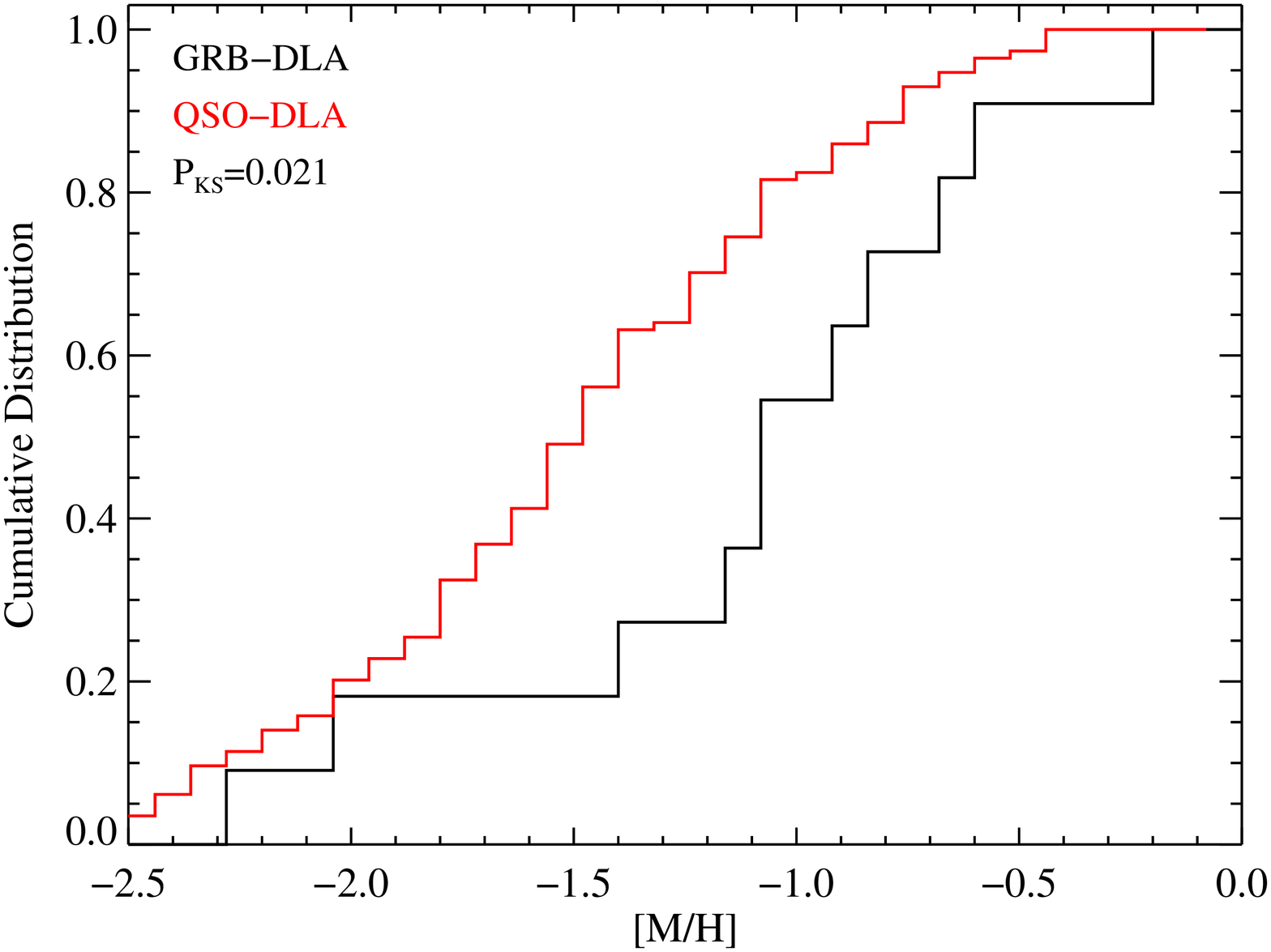} 
\caption{
The histograms show the cumulative distribution of QSO-DLA and LGRB-DLA
metallicities in the statistical samples compiled by \cite{pgw+03} and
Prochaska et al. (2007). As seen, the LGRB-DLA metallicities are systematically
higher than the QSO-DLA metallicities. See also Fynbo et al. (2008) for
a simple model of these two distributions.
}
\label{Z}
\end{center}
\end{figure}

Afterglow spectroscopy often allows us to measure the
metallicity of the line-of-sight in the host galaxy. In Fig.~\ref{Z} we plot the
metallicities along LGRB sightlines together with metallicities derived from QSO
damped Lyman-$\alpha$ absorbers (QSO-DLAs). Here it can be seen that LGRBs are
more metal rich than QSO-DLAs at similar redshifts. Some of the LGRB sightlines
are almost as metal rich as the Lyman-break galaxies at similar redshifts 
(Pettini et al. 2001). The shift in metallicity relative to QSO-DLAs can be
understood from the different selection functions of the (star-formation
selected) LGRB-DLAs and the (HI cross-section selected) QSO-DLAs combined
most likely with metallicity gradients in high-z galaxies (Fynbo et al.
2008). Hence, most likely LGRBs will give a reasonably unbiased census of
where the massive stars are located, at least at $z > 2$ (Fynbo et al. 2006b). 

\subsection{Extinction}
In addition to HI column densities, metal and molecular abundances and
kinematics, the afterglow spectra also provide information of the extinction
curves. The intrinsic spectrum of the afterglow is from theory predicted to be
a power-law and therefore any curvature or other broad features in the spectrum
can be interpreted as being due to features in the extinction curve. So far,
almost all the extinction curves derived for LGRB host galaxy sightlines have
been consistent with an extinction curve similar to that of the SMC.  Recently,
we obtained the clearest yet detection of the 2175 \AA \ bump known from the
Milky Way in a z=2.45 LGRB (Eliasdottir et al.\ 2008 and
Fig.~\ref{bump}).  This LGRB absorber also has unusually strong metal lines
suggesting that the presence of the 2175 {\AA}  extinction bump is related to a
high metallicity.  However, we have examples of LGRBs with nearly solar
metallicity for which the bump is not seen so it seems that metallicity is not
the only parameter controlling the presence of the 2175 {\AA} extinction bump.
Concerning the amount of extinction the LGRB sightlines vary from no extinction
(e.g., GRB050908, Fig.~\ref{LyC}) to $A_V > 5$ mag (e.g., GRB070306, Jaunsen et
al. 2008).

\section{Conclusions and outlook}
Spectroscopy of LGRB afterglows provides redshifts and information on the ISM
properties for the population of galaxies containing the bulk of
the high-redshift massive stars. We are currently working on securing this
information for a complete sample of Swift LGRBs. 
Already now we can conclude that the environments
of LGRB progenitors are very diverse with metallicities ranging from solar to
a hundredth solar and extinction ranging from none to A$_V>5$ mag. We have also
identified a sightline with significant escape of Lyman continuum photons
and another with a clear 2175 \AA \ extinction bump. 

Even though the completeness of the current {\it Swift} sample in terms
of detections of optical afterglows ($\sim$75\%), redshift determinations 
($\sim$50\%) and host galaxy detections ($\sim$80\%) is much higher than 
for LGRBs from previous missions we still need to do better - preferably 
all three fractions should be $\gtrsim80$\%. We also need to improve the
understanding of the link between LGRBs and massive stars. The SN-less
GRBs GRB\,060505 and GRB\,060614 show that either some LGRBs are unrelated
to massive stellar death or some massive stars die without causing SNe.
The currently unfolding SN2008D/XRF080109 shows that there may be much
more frequent bursts with softer prompt emission bridging the gap (in terms
of both burst, SN and host properties) between
GRBs and normal Ic SNe (e.g., Soderberg et al.\ 2008; Malesani et al.\ 2008).

\acknowledgements
We thank our collaborators and
the {\it Swift} team for carrying out such a wonderful experiment.  The Dark
Cosmology Centre is funded by the DNRF.

\begin{discussion}

\discuss{Burbidge}{ 
Some closeby -low redshift- GRBs are clearly associated with supernovae-
we see traces of the SN light curves.
But for most of the GRBs either no optical or radio object has been
identified or only very faint afterglows are found.
Thus the connection with galaxies- even star forming galaxies is much
weaker. On the other hand, many optical spectra
of GRB afterglows show the absorption features of QSOs. Thus I think
that some attention should be paid to the connection of GRBs with QSOs.
Here the MgII absorption seen in all of the afterglows with absorption
but only in a fraction of QSOs presents a real problem
for those who believe that the QSOs all have cosmological redshifts.
}

\discuss{Fynbo}{
I would describe the situation as follows. For $z<2$ GRBs we almost always 
detect a host galaxy at the position of the afterglows. For more 
distant GRBs the fraction of hosts detected to a detection limit of
about $R=27$ drops. The properties of the GRB absorption systems are
most similar to the DLAs seen in QSO spectra although GRB systems on 
average are more metal rich. In no case has a GRB been associated directly
with a galaxy hosting an AGN.
}

\discuss{Rauw}{
There have been suggestions to use nuclear resonance absorption lines in
the $\gamma$-ray spectrum to determine the redshifts of GRBs beyound 
z=6. That would require very large column densities, well in excess of the 
values you have presented. Can you comment on this?}

\discuss{Fynbo}{
I am affraid I do not have much insight into this issue. I can add that 
in some cases substantially higher column densities are inferred from X-ray 
absorption than from Hydrogen Lyman-$\alpha$ (e.g., Watson et al. 2007,
ApJL, 660, L101). This is most likely due to ionized material close to
the GRB progenitor. Still, the column densities inferred from X-ray
absorption are so far all below 10$^{23}$ cm$^{-2}$ equivalent HI.}

\end{discussion}

\end{document}